# Probing the ultimate limits of metal plasticity

Luis A. Zepeda-Ruiz[1], Alexander Stukowski[2], Tomas Oppelstrup[1], and Vasily V. Bulatov[1]

[1]Lawrence Livermore National Laboratory, Livermore, California, USA
[2]Technische Universität Darmstadt, Darmstadt, Germany

**Along with high strength, plasticity is what makes metals so widely usable in our material world. Both strength and plasticity properties of a metal are defined by the motion of dislocations - line defects in the crystal lattice that divide areas of atomic planes displaced relative to each other by an interatomic distance. Here we present first fully dynamic atomistic simulations of single crystal plasticity in metal tantalum predicting that above certain maximum rate of straining - the ultimate limit - the dislocations can no longer relieve mechanical loads and another mechanism, twinning, comes into play and takes over as the dominant mode of dynamic response. At straining rates below the ultimate limit, the metal attains a path-independent stationary state of plastic flow in which both flow stress and dislocation density remain constant indefinitely for as long as the straining conditions remain unchanged. In this distinct state tantalum flows like a viscous fluid while still remaining a strong and stiff metal.**

Plasticity – permanent change of material shape - results from dislocation lines sweeping through the lattice planes, each line propagating a tiny quantum of atomic displacement - the Burgers vector - and thus relieving lattice strain. Strength of a plastic metal is defined by the magnitude of mechanical force (load) needed to push dislocations through the lattice. When dislocation motion alone is insufficient to relieve lattice strain, metal responds to straining in some other, possibly catastrophic manner, e.g. twinning or fracture. Our key objective here is to answer the following question: under what conditions of straining does dislocations' innate ability to relieve stress becomes overwhelmed?

The central question posed above appears to be squarely within the purview of Dislocation Dynamics (DD), a method developed specifically for simulations of crystal plasticity resulting from dislocation motion[3,4]. However application of DD for our stated purpose here is problematic due to the method's own limitations key of which is that DD models do not include any other mechanisms of inelastic response alternative to dislocation motion, e.g. fracture or twinning. Because a DD model cannot decide when any such behaviors become active it does not know when dislocation plasticity reaches its limits. Here we employ Molecular Dynamics (MD) simulations to probe the limits of metal plasticity. Our MD simulations incorporate every "jiggle and wiggle"[5] of atomic dynamics and are thus faithful to every possible mechanism of dynamic response.

MD simulations of bulk crystal plasticity are computationally challenging and, to the best of our knowledge, have not been carried out before. In addition to the notorious length and time scale limits (that the less expensive DD method was developed to overcome), another serious challenge is that MD simulations generate enormous amounts of data. An MD simulation of the kind reported below generates about 1 exabyte (=$10^{18}$ bytes) of digital data in just one day on the full Sequoia supercomputer[6], an amount comparable to Google's estimated worldwide storage capacity. To be useable, MD data must be compressed to size and recast into a form that a human can grasp and comprehend. Here we rely on recently developed methods of *in situ* computational microscopy to identify and precisely characterize extended defects in crystal[7,8] and reduce the amount of data to deal with by orders of magnitude (for more details see Methods).

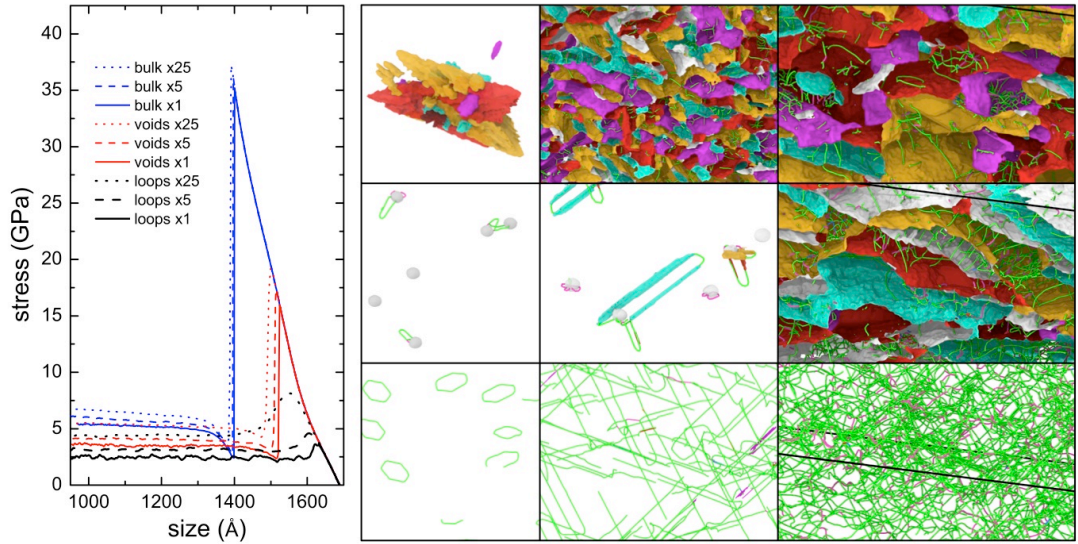

**Figure 1:** (Left) Stress computed in MD simulations of uniaxial compression along [001] axis as a function of the logarithm of the specimen size. Blue, red and black curves are computed for the initially perfect crystal, crystal with voids and crystal with dislocations, respectively. Solid, dashed and dotted lines correspond to straining rates x1, x5 and x25, respectively. (A) Three upper row snapshots depict co-nucleation of embryonic twins (left), twin propagation (middle) and twin growth (right) in the initially perfect crystal. (B) The middle row snapshots depict dislocation nucleation at voids (left), nucleation of embryonic twins on stretched screw dislocations (middle) and twin growth (right) in the crystal with initial voids. (C) The bottom row snapshots depict initial dislocation sources (left), extension of dislocations from the sources and their impingement (middle), and formation of a dense dislocation network in the crystal with initial dislocation sources (right). Most dislocations appear as green lines with some magenta lines present. The twins appear as hollow volumes bounded by interfaces colored red, yellow, purple and green to distinguish four distinct rotational twin variants. The volumes of the parent (untwined) crystal are bounded by the light gray interfaces.

In this Letter we focus on the dynamic response of tantalum single crystal to compressive straining along the [001] crystal axis. In our MD simulations the metal is subjected to simple and fully controlled straining conditions in which pressure, temperature and (true) straining rate are all kept constant (see Methods for more detail). First we compare the response to compression of tantalum crystal starting from three different initial configurations: (A) defect-free perfect crystal, (B) the same crystal with 24 randomly placed voids[1] and (C) 24 dislocation loops placed into the same locations as voids in B. Shown in Fig. 1 are the stress-strain curves computed from these three starting configurations under three different straining rates: $\dot{\varepsilon}_T = 1.109 \cdot 10^7$ 1/s

(our minimal or "base" rate denoted x1), $\dot{\varepsilon}_T = 5.545\ 10^7$ 1/s (x5) and
$\dot{\varepsilon}_T = 2.773\ 10^8$ 1/s (x25). Response of three crystals to straining is markedly different. Crystals A and B give in (yield) to strain abruptly after reaching peak stress of 37 GPA (crystal A) and 19 GPa (crystal B) followed by sharp drops in stress, whereas crystal C yields gradually along smooth stress-strain curves passing over peak stresses that are both much lower and distinctly rate-dependent. The latter response is comparable to that of ductile metals observed in laboratory experiments[9].

Analysis of crystal configurations attained in these simulations reveals that crystals A and B yield by twinning - sudden strain-induced reorientation of the crystal lattice within bounded patches of the material[10] (Fig. 1(A) and 1(B)). In contrast, crystal C does not twin but yields through rapid motion and multiplication of the initial dislocations forming dense dislocation networks during and soon after yielding (Fig. 1(C)). This comparison shows that, by depriving the metal of dislocations in crystals A and B, not only is the yield strength over-predicted by a large factor, but the metal yields through a mechanism qualitatively different from crystal C in which dislocations are introduced from the beginning. As is well known, *multiplication* of pre-existing dislocations entails much lower stress that dislocation *nucleation* whatever its detailed mechanism[11]. As no ductile metal has ever been obtained dislocation-free, stress-strain behaviors observed in crystals A and B are perhaps unrealistic. In the following we confine ourselves to simulations performed on crystals with pre-existing dislocations.

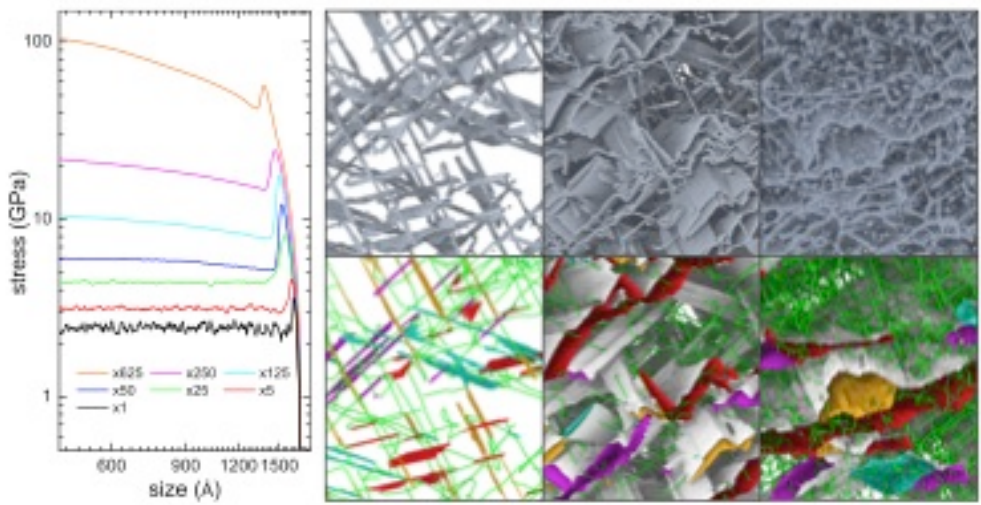

**Figure 2:** (Left) Log-log plot of stress as a function of specimen size computed in MD simulations of specimen compression at constant rates. The curves are colored according to the compression rates shown at the bottom. (Right) Snapshots extracted from simulations at rate x50 and depicting twin nucleation on screw dislocations, twin propagation and growth, from left to right. The upper three snapshots show only the atoms with highly distorted local coordination (these naturally depict crystal defects), the lower three snapshots show the same configurations processed using methods of *in situ* computational microscopy (see Methods for details). Coloring of lines and interfaces in the snapshots is the same as in figure 1.

We define an ultimate limit of metal plasticity as the maximum straining rate $\dot{\varepsilon}_M$ that the material can sustain on dislocation motion alone, without triggering some other mode of inelastic response – e.g. twinning or fracture. From the data shown in Fig. 1 it is clear that $\dot{\varepsilon}_M \geq x25$. To continue probing for $\dot{\varepsilon}_M$ we next ran MD simulations in which crystal C was subjected to still higher straining rates. As shown in Fig 2(a), both the upper yield stress (stress maximum) and the flow stress rise sharply with the increasing straining rate prompting us to use the log scale for stress on the figure. Response to straining is markedly different at rates above and below x25: whereas at rates ≤x25 the crystal quickly settles down at a flow stress that remains unchanged through the rest of the straining simulation, at rates >x25 the flow stress steadily increases after the post-yield drop - the metal is hardening[12]. Analysis of atomic snapshots reveals that at rates >x25 the crystal twins profusely whereas no twinning is observed at rates ≤x25. Although tempted to conclude that $x25 \leq \dot{\varepsilon}_M \leq x50$, we note that in all simulations at rates >x25 twinning is triggered during an initial yield transient that depends on the starting defect configuration (microstructure). We further report, in agreement with numerous experimental observations [ref], that under the same straining rate yield response depends on the density of the initial dislocation lines. In particular, scarcity of pre-existing dislocations results in a high yield stress, plentiful twinning and noticeable hardening even at rate x25 previously deemed to be below $\dot{\varepsilon}_M$ (see Supplementary Information for more details).

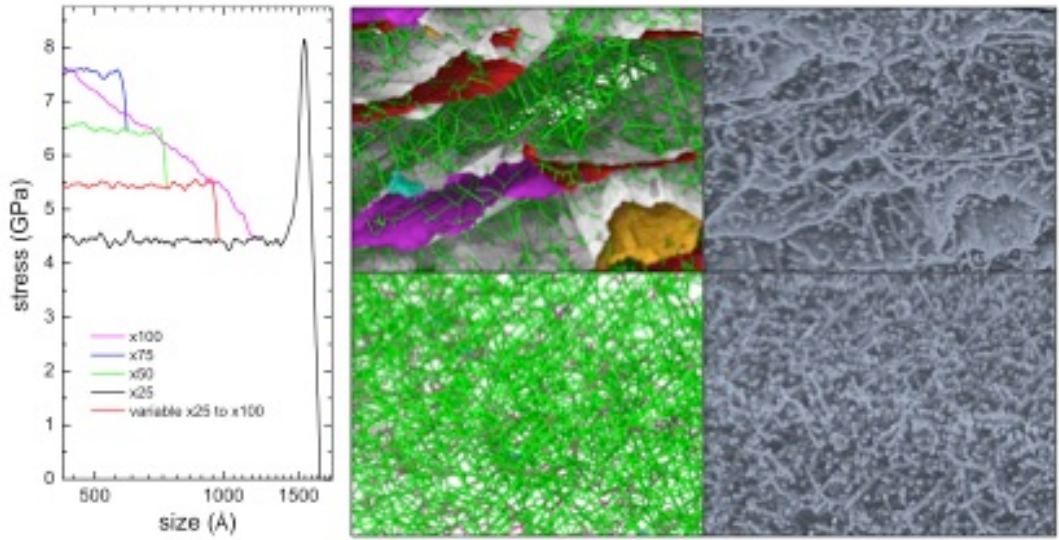

**Figure 3:** (a) Stress-strain response computed in MD simulations of tantalum compressed under constant true straining rates following strain rate jumps and under a variable rate linearly ramped up from x25 to x100. Each curve is labeled with the straining rate at which it was computed. (b) The upper snapshots are two renditions of the same heavily twinned configuration attained under straining at rate x50 starting from the initial (unstrained) configuration. The upper snapshots are from the simulation in which rate x50 was attained after pre-straining at rate x25. Color coding is the same as in figures 1 and 2.

To make our definition of $\dot{\varepsilon}_M$ more precise, we intend to continue to probe for this limit by ramping the straining rate up gradually, either continuously or in steps, thus giving dislocations ample (ideally infinite) time to multiply and to avoid the stress overshoot observed at initial yielding. First, starting from a state attained soon after the flow stress settles in the simulation at rate x25 we suddenly increase the straining rate to x50 and continue at this rate to allow the crystal to settle at a new flow stress (Fig. 3). Then the rate is suddenly raised to x75 and, after some more straining at this rate, to x100. After each step up in the straining rate, we observe the stress to quickly settle to a new higher value while showing no discernible yield overshoot. However, while no twins were detected at x25, x50 and x75, under the highest rate x100 the crystal develops small but detectable twins thus bracketing the ultimate limit down to $x75 < \dot{\varepsilon}_M < x100$. To verify this estimate we performed yet another simulation in which the straining rate increased linearly $\dot{\varepsilon}(t) = \dot{\varepsilon}_1 + \ddot{\varepsilon} t$ starting from the same state at $\dot{\varepsilon}_1 = x25$. The ramping rate $\ddot{\varepsilon} = -5.545 \cdot 10^{17} s^{-2}$ was selected so as to reach rate x100 in the very end when the specimen becomes fully compressed. Stress response to ramped compression is also shown in Fig. 3. Careful analysis reveals that first embryonic twins form when the straining rate reaches x80, or $\dot{\varepsilon}_M \approx 8 \cdot 10^{9}$ 1/s. Within our limited computational resources and to within the accuracy of our model of interatomic interaction, the latter figure is our best estimate for the ultimate rate that tantalum metal can sustain on dislocation plasticity alone under compression along its [001] axis.

Our simulations reveal tight interplay between dislocation plasticity and twinning. On one hand, dislocation motion relieves stress and prevents the onset of twinning. On the other hand, when dislocations are unable to keep up with rising stress, they instigate twinning serving as preferential sites for heterogeneous nucleation of twins via the mechanism proposed by Sleeswyk's[13,14]. In the presence of dislocations, the threshold stress for twinning is estimated at ~7 GPa whereas the same threshold in a perfect crystal is about five times higher. In the twinned crystals dislocations appear to be less numerous in the twin interiors than in the (un-twinned) parent crystal (Fig. 2). We speculate that, once nucleated, the twins grow preferentially into the regions in the crystal that are relatively devoid of dislocations and/or that twin boundaries block dislocations from moving inside the twins. That the latter factor is at play is also suggested by hardening - a gradual increase in the flow stress with the increasing strain - the more prominent the higher the twin fraction. In contrast, in all simulations where twinning does not occur, we observe no hardening: once it settles down past the initial yield transient, the flow stress remains manifestly constant down to the very end of compressive straining.

We posit that, if brought sufficiently slowly to a constant true straining rate below $\dot{\varepsilon}_M$, single crystal tantalum attains and sustains a stationary state of plastic flow for as long as temperature, pressure and straining rate are maintained constant. Our simulations thus far substantiate this assertion: all simulations brought to a constant rate at and below x80 - instantly, in steps or via continuous ramping - reach a steady flow stress after a short initial yield transient and stay at this stress to the very end of each compression simulation (Fig. 3). To further validate the existence of a distinct state of plastic flow we performed several simulations in which the crystal was first pre-strained past yield at a high rate followed by still further straining at a lower rate. As seen in Fig. 4(a), the flow stress is manifestly independent of the straining path and is the same no matter how the rate was attained - applied from the beginning, or reached by stepping the rate up from below or by dialing the rate down from above. While fully retaining its crystal lattice and remaining strong and elastically stiff, tantalum metal responds to straining just like a perfectly viscous fluid.

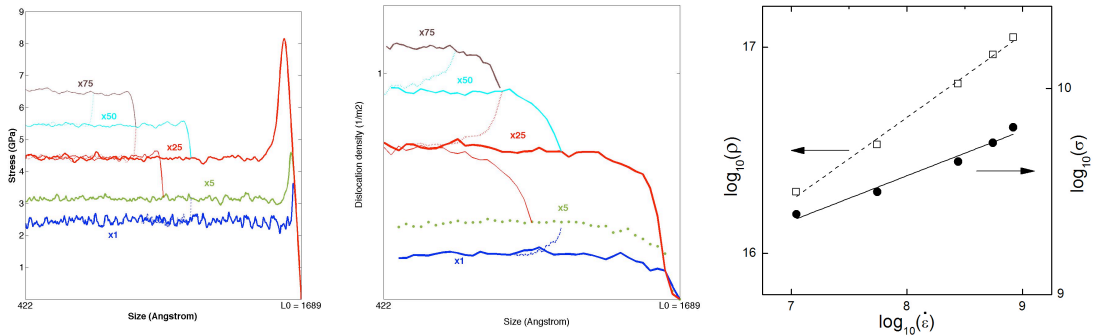

**Fig. 4:** (a) Stress-strain response to straining rate jumps plotted in the semi-log coordinates: the lines are directly labeled with the straining rates at which they were computed. Curves computed for the same rate have the same colors. (b) Dislocation density-strain curves from the same simulations. Line labeling is the same as in (a). (c) Log-log plot of the steady state dislocation density (in m$^{-2}$) and the flow stress (in Pa) as functions of the straining rate (in 1/s). The lines show linear fits approximating the data as $\rho \sim \dot{\varepsilon}^{0.4}$ and $\sigma \sim \dot{\varepsilon}^{0.2}$.

Additional evidence for the attainment of a stationary flow state is the behavior of dislocation line density $\rho_t$ during straining. Defined as the total length of dislocation lines in a unit volume, dislocation density is seen to attain a stationary value of its own, even if it takes markedly longer time to settle compared to the flow stress (Fig. 4(b)). Furthermore, just like the flow stress, the steady value of the dislocation density is path-independent and is the same, within statistical errors, no matter how the rate was attained. We regard the observed state of flow as a distinctive non-equilibrium stationary state maintained at a constant rate of energy dissipation equal to the product of the flow stress and the straining rate. In this stationary open system, mechanical energy is supplied by straining, then converted though dislocation motion into heat and eventually collected and removed by the thermostat. In a further confirmation of its stationary nature, we show that flow stress remains unchanged over arbitrarily large compressive strains in a "metal kneading" simulation detailed in Supplementary Information.

Within the range of one order of magnitude of saturated (stationary) dislocation densities observed in our simulations, the flow stress appears to scale approximately as

$\sigma \sim \sqrt{\rho}$ (Fig. 4(c)) which is expected when the resistance to dislocation motion is defined by dislocation intersections (forest hardening)[15,16]. The dislocations appear to move in a stop-and-go fashion remaining relatively motionless against obstacles (junctions and self-pinning cusps) for periods of time intermittent with bursts of faster motion. The distribution of dislocation line orientations is non-uniform dominated by line segments nearly parallel to their Burgers vectors (screws). Average dislocation velocities estimated from Orowan's equation $\dot{\varepsilon}_p = \rho b v$ are surprisingly low ranging between 3 and 40 m/s, however dislocations are seen to move considerably faster right after jumping an obstacle (see Supplementary Information). Along with numerous dislocations, we observe copious amounts of debris - vacancies, interstitials and defect clusters - produced by dragging jogs (scars) left on the screw dislocations after intersections with other dislocations and generated as cross-kinks (see Supplementary Information)[14,17].

To our knowledge, no experimental evidence exists for a stationary state of plastic flow in metals. Obviously, direct comparisons to quasi-static ($\dot{\varepsilon} \sim 10^{-3}$ 1/s) straining experiments are risky since our straining rates are about 10 orders of magnitude higher. At the same time, dynamic simulations like ours allow one to exercise full control over straining conditions, which is something straining experiments always strive for but rarely achieve. In particular, over the range of strains as high as in our simulations, grip friction often leads to rotation of the straining axis and macroscopic distortions in the specimen shape such as barreling under compression[18] or failure by necking and fracture under tension[19]. Furthermore, the experimental practice to maintain a constant machine grip velocity - as opposed to constant true straining rates employed in our simulations - can mask the true plastic response beyond recognition[20]. For example, when performed under a constant engineering rate (= constant grip velocity) our compression simulations reveal marked hardening with flow stress increasing by as much as 100% (Fig. 3). We maintain that - at least in our model of tantalum - this hardening is only apparent, owing to a four-fold decrease in the specimen dimension in the end of compression simulation and, correspondingly, a four-fold increase in the true straining rate that drives the flow stress higher.

**METHODS**
**Molecular Dynamics simulations**. The simulations were performed using the open-source code LAMMPS[2] on a rectangular fragment of tantalum crystal with dimensions $L_x \times L_y \times L_z = 128 a_0 \times 256 a_0 \times 512 a_0$ oriented along the principal [100], [010], [001] axes of the cubic lattice, respectively (here $a_0 = 0.33$ nm is the lattice constant of the body-centered cubic lattice of tantalum). Interaction among tantalum atoms was modeled with the EAM interatomic potential developed by Li et al.[21] The number of atoms in the simulation box was approximately 33 million and 3D periodic boundary conditions were applied to embed the box seamlessly into an infinite crystal. The crystal was compressed either at a constant or a variable "true" rate $\dot{\varepsilon}_T = \frac{1}{L_z}\frac{dL_z}{dt}$ along its initially longest dimension $L_z$ while the two lateral dimensions $L_x$ and $L_y$ were left to dynamically expand to maintain $xx$ and $yy$ components of the mechanical stress near zero. Most straining simulations were continued until the "longest" dimension $L_z$ was compressed to 1/4 of its initial size by which time $L_x$ and $L_y$ almost doubled thus keeping volume and pressure nearly constant throughout the simulation. Temperature was also maintained near 300°K using a Langevin thermostat (see Supplementary Information for more simulation details). Everywhere in the text by "stress" we mean the uniaxial compressive stress $\sigma_{zz}$ that develops in the model crystal in response to straining along its [001] z-axis.

***In situ* computational microscopy.** In addition to the standard local atom filtering techniques[7] we employ the dislocation extraction (DXA) and the grain segmentation (GSA) algorithms to reveal structural defects in simulated crystals and to reduce the amount of data to deal with by orders of magnitude. The DXA algorithm accurately traces and indexes the dislocation lines and builds a concise network representation of crystal microstructure identical to the one employed in the DD method[8]. The newly developed GSA algorithm automatically partitions the atomistic crystal into regions of similar crystal orientation (grains) and generates a geometric representation of the interfaces delineating the grains. This conversion from the fully atomistic model to a highly reduced description of the defect microstructure is performed at regular time intervals during the MD simulation and enables us to follow the motion and reactions of dislocations, the onset of twinning and other important events in arbitrary degree of detail (see Supplementary Information for additional examples of *in situ* computational microscopy).

**Supplementary Information** is linked to the online version of the paper at www.nature.com/nature.

**Acknowledgements** This work was performed under the auspices of the US Department of Energy by Lawrence Livermore National Laboratory under Contract W-7405-Eng-48. This work was supported by the NNSA ASC program. Computing support came from the Lawrence Livermore National Laboratory (LLNL) Institutional Computing Grand Challenge program. We thank W. Cai, B. Sadigh, N. Barton, A. Arsenlis and W. Brett for helpful suggestions.

**Author Information** Reprints and permissions information is available at npg.nature.com/reprintsandpermissions. The authors declare no competing financial interests. Correspondence and requests for materials should be addressed to VVB (bulatov1@llnl.gov).